\documentclass[a4paper,12pt]{article}

\usepackage{amssymb,amsfonts,amsmath,epsfig,float}

\usepackage{mathrsfs}

\newcommand{\be}{\begin{eqnarray}}
\newcommand{\ee}{\end{eqnarray}}
\newcommand{\nnee}{\nonumber\ee}

\newcommand{\upd}{{\rm \,d}}

\newtheorem{theorem}{Theorem}[section]

\newtheorem{proposition}[theorem]{Proposition}
\newtheorem{definition}[theorem]{Definition}
\newtheorem{notation}[theorem]{Notation}
\def\beginproof{\par\strut\vskip 0.1cm\noindent{\bf Proof}\par}
\def\endproof{\par\strut\hfill$\square$\par\vskip 0.5cm}

\newcommand{\Tr}{\,{\rm Tr}\,}
\def\Mo{{\mathbb M}}
\def\Io{{\mathbb I}}

\def\tinyK{{\mbox{\tiny K}}}
\def\sanull{{\cal A}_{\mbox{\tiny sa}}^0}

\def\HS{{\mbox{\tiny HS}}}

\title{Parameter-free description of the manifold of non-degenerate density matrices} 

\author{Jan Naudts\\
Universiteit Antwerpen\\
\small Physics Department, Universiteitsplein 1, 2610 Antwerpen, Belgium\\
\small	\url{Jan.Naudts@uantwerpen.be}\\
\small	\url{https://orcid.org/0000-0002-4646-1190}
}
\date{}

\begin{document}

\maketitle

\begin{abstract}
The paper gives a definition of exponential arcs in the manifold of non-degenerate
density matrices and uses it as a starting point to develop a parameter-free version
of non-commutative Information Geometry in the finite-dimensional case.
Given the Bogoliubov metric the m- and e-connections are each other dual. 
Convex potentials are introduced. They allow to introduce dual charts.
Affine coordinates are introduced at the end to make the connection
with the more usual approach.
\end{abstract}


\section{Introduction}

Models belonging to the quantum exponential family have been intensively studied
within Statistical Physics long before Amari \cite{AS85,AN00} introduced
dually-flat geometries into the theory of statistical models.
The generalization of Amari's work to quantum models was taken up by Hasegawa and others
\cite{HH93,PS96,OSA96,HH97,HP97}.
See for instance Chapter 7 of \cite{AN00} and the book of Petz \cite{PD08}.

In a series of papers \cite{PS95,GP98,PR99,PC07,PG13, PG13b}
Pistone and coworkers developed a parameter-free approach to Information Geometry.
Similar efforts are found in the work of Newton \cite{NN12,NN19}.

In the book of Ay et al \cite{AJVLS17}, Chapter 3.3,
two distinct approaches are mentioned. In Pistone's approach the manifold of probability 
measures compatible with a given measure receives the structure of a Banach manifold.
Alternatively, a manifold of probability measures can receive its geometry from
its embedding in a linear space of signed measures.

Early efforts to generalize Pistones approach to the quantum context include
the works of Grasselli and Streater \cite{GS01,SRF04a,SRF04b,GMR10} and 
of Jen\v cov\'a \cite{JA06}.
Recently, a different line of research is started by Ciaglia et al \cite{CIJM19}.
They study the action of the group of invertible operators on the manifold
of density operators.

Technical problems appear when considering continuous measure spaces.
Such problems are avoided here 
by restriction to the finite-dimensional case.

From the Literature the following guidelines are adopted.
\begin{description}
 \item [\,-\quad] The manifold has a maximal extent; models belonging to an exponential family
 describe submanifolds;
 \item [\,-\quad] The manifolds are Banach manifolds; charts take values in a Banach space;
 \item [\,-\quad] Each point of the manifold is the centre of a chart;
 \item [\,-\quad] The geodesics are exponential arcs;
 \item [\,-\quad] The metric can be obtained from a divergence function by differentiation;
 \item [\,-\quad] Parallel transport is used to derive the geometric connection.
\end{description}

In Quantum Information Theory \cite{PD08} Bures' distance \cite{BD69,UA76}
is extensively used. It is the quantum analogue of the Hellinger
distance and has quite unique properties. Never the less the
inner product needed in the present context is that of Bogoliubov
\cite{NVW75,PT93,GS01, NJ18}. For this inner product the e- (exponential) and m- (mixture) connections
become each other duals \cite{GS01}. A proof follows below in Section \ref {sect:dualgeo}.

Let me finally point out that the generalization of Information Geometry
to the non-commutative context is characterized by non-uniqueness.
Section 10.3 of \cite{PD08} discusses a class of metrics that all
generalize the Fisher information metric to the quantum context.
In addition, the notion of exponential arcs, which is the topic
of the present work, is non-unique. An alternative definition 
given in \cite{NJ20} introduces exponential arcs of faithful
states on a $\sigma$-finite von Neumann algebra.

The next section gives the definition of exponential arcs of density matrices.
Vectors tangent to these arcs are discussed in Section \ref {sect:tang}. 
The exponential map is shown to be well-defined, one-to-one and onto.
A chart affine for the e-connection is discussed. 
In Sections \ref {sect:met} and \ref {sect:dualgeo} Bogoliubov's inner
product is introduced. Parallel transport is used to derive
the covariant derivative and its dual.
Sections \ref {sect:leg} and \ref {sect:dualchart} introduce convex potentials
and dual charts.
The link with parameterized approaches is made in Section \ref {sect:aff}.
At the end follows a section with summary and discussion.

\section{Exponential arcs}
\label{sect:arc}

A first step in the construction of a geometry on the manifold $\Mo$ of non-degenerate
density matrices of dimension $n$-by-$n$ is the choice of
the geodesics that will be used to connect pairs of points in the manifold.

The following definition generalizes the concept introduced by Cena and Pistone \cite{PC07,PG13}
to the non-commutative context.

\begin{definition}
An {\em exponential arc} connecting the density matrix $\sigma$
to the density matrix $\rho$ is a map $t\mapsto\sigma_t$
with $\sigma_t$ given by
\be
\sigma_t=\exp((1-t)\log\rho+t\log \sigma-\alpha(t))
\nnee
and with $\alpha(t)$ given by
\be
\alpha(t)&=&\log\Tr\exp((1-t)\log\rho+t\log\sigma).
\nnee
\end{definition}

Note that, given $\rho$ and $\sigma$ in $\Mo$, $\sigma_t$ with $t\in[0,1]$ is a non-degenerate density matrix belonging to the
manifold $\Mo$. It satisfies $\sigma_0=\rho$ and $\sigma_1=\sigma$. The normalization function
$\alpha$ satisfies $\alpha(0)=\alpha(1)=0$.

For further use introduce the following notation.

\begin{notation}
For any pair of density matrices $\rho$ and $\sigma$ in $\Mo$ the tangent vector $Y_\rho(\sigma)$
is given by
\be
Y_\rho(\sigma)&=&\frac{\upd\,}{\upd t}\bigg|_{t=0}\sigma_t,
\nnee
where $t\mapsto\sigma_t$ is the exponential arc connecting $\sigma$ to $\rho$.
\end{notation}

Use the identity (see \cite{AN00} p.~156 for a proof)
\be
\frac{\upd\,}{\upd t}\bigg|_{t=0}e^{H+tA}
=
\int_0^1\upd u\,e^{uH}Ae^{(1-u)H}
=
\int_0^1\upd u\,e^{(1-u)H}Ae^{uH}
\label{bogo:ident}
\ee
to calculate
\be
Y_\rho(\sigma)
&=&
\int_0^1\upd u\,\rho^u\left[\log\sigma-\log\rho
-\frac{\upd\,}{\upd t}\alpha(t)\bigg|_{t=0}\right]
\rho^{1-u}.
\nnee
Note that
\be
\frac{\upd\,}{\upd t}\alpha(t)\bigg|_{t=0}
&=&\Tr\int_0^1\upd u\,\rho^u[\log\sigma-\log\rho]\rho^{1-u}\cr
&=&-D(\rho||\sigma)
\nnee
with $D(\rho||\sigma)$ Umegaki's relative entropy \cite{UH62}
\be
D(\rho||\sigma)&=&\Tr \rho(\log \rho-\log\sigma).
\label{2:relent:def}
\ee

\begin{notation}
\label{not:special}
Each matrix $A$ 
defines a matrix denoted $[A]^\tinyK_\rho$ by the relation
\be
[A]^\tinyK_\rho&=&\int_0^1\upd u\,\rho^u \,A(\rho)\,\rho^{1-u},
\qquad \rho\in\Mo.
\nnee
\end{notation}
The map $A\mapsto [A]^\tinyK_\rho$ is the Kubo transform \cite{PD08}.

\begin{notation}
Given a pair of density matrices $\rho$ and $\sigma$ let
\be
c_\rho(\sigma)
&=&
\log\sigma-\log\rho+D(\rho||\sigma).
\ee
\end{notation}

Note that $\Tr\rho\, c_\rho(\sigma)=0$.
With these notations one can write the tangent vector as 
\be
Y_\rho(\sigma)
&=&
\left[c_\rho(\sigma)\right]^\tinyK_\rho.
\label{exp:tanvec}
\ee

\section{The tangent plane}
\label{sect:tang}

Fix a non-degenerate density matrix $\rho$ in $\Mo$.
The tangent plane $T_\rho\Mo$
at the point $\rho$ in the manifold $\Mo$ is the space of derivatives 
at the origin $t=0$ of exponential arcs $t\mapsto\sigma_t$
connecting any density matrix $\sigma$ in $\Mo$ to the density matrix $\rho$.
Let us characterize this space.

\begin{proposition}
\label{tan:lem}
For any $n$-by-$n$ matrix $V$ with vanishing trace 
there exists an $n$-by-$n$ matrix $A$ such that
$V=[A]^\tinyK_\rho$.
If $V$ is Hermitian then $A$ is Hermitian as well.
If in addition, the trace of $V$ vanishes then the expectation $\Tr\rho A$ vanishes as well.
\end{proposition}

\beginproof
Consider an orthonormal basis $(e_i)_i$ in which $\rho$ is diagonal.
One has $\rho e_i=\lambda_i$ with $\lambda_i>0$. The matrix $A$ with matrix
elements given by
\be
\langle e_j|Ae_i\rangle&=&\frac{\langle e_j|Ve_i\rangle}{\int_0^1\upd u\,\lambda_i^u\lambda_j^{1-u}}
\nnee
satisfies the requirements.

If $V$ is Hermitian then one has
\be
\langle e_j|Ae_i\rangle\,\int_0^1\upd u\,\lambda_i^u\lambda_j^{1-u}\,
&=&\langle e_j|Ve_i\rangle=\overline{\langle e_i|Ve_j\rangle}\cr
&=&
\overline{\langle e_i|Ae_j\rangle}\int_0^1\upd u\,\lambda_i^u\lambda_j^{1-u}.
\nnee
This shows that also $A$ is Hermitian.

If in addition $\Tr V=0$. Then one has
\be
\Tr\rho A
&=&
\sum_i\lambda_i\langle e_i|Ae_i\rangle\cr
&=&
\sum_i\lambda _i\frac{\langle e_i|Ve_i\rangle}{\lambda_i}\cr
&=&
\Tr V\cr
&=&
0.
\nnee
\endproof

For convenience, the following notations are introduced.
\begin{notation}
The linear space of Hermitian matrices $V$ with vanishing trace $\Tr A=0$ is denoted $\sanull$. 
The linear space of Hermitian matrices $A$ with vanishing expectation $\Tr\rho A=0$ 
given $\rho$ in $\Mo$ is denoted ${\cal A}_\rho$.
\end{notation}

\begin{proposition}
The tangent space $T_\rho\Mo$ consists of all Hermitian $n$-by-$n$ matrices
with vanishing trace: $T_\rho\Mo=\sanull$.
\end{proposition}

\beginproof
Let $V$ be any matrix in $\sanull$.
By the previous proposition there exists a Hermitian $n$-by-$n$ matrix 
$A$ such that $V=[A]^\tinyK_\rho$ holds.
Let $\sigma$ be defined by
\be
\sigma&=&\frac{\exp(\log\rho+A)}{\Tr \exp(\log\rho+A)}.
\nnee
Then $\sigma$ is a density matrix.
Let $t\mapsto\sigma_t$ denote the exponential arc connecting $\sigma$ to $\rho$.
The tangent vector at $t=0$ is given by
\be
Y_\rho(\sigma)
&=&
\int_0^1\upd u\,\rho^u\left[\log\sigma-\log\rho
-\frac{\upd\,}{\upd t}\alpha(t)\bigg|_{t=0}\right]
\rho^{1-u}\cr
&=&
\int_0^1\upd u\,\rho^u\left[A-\log\Tr \exp(\log\rho+A)
-\frac{\upd\,}{\upd t}\alpha(t)\bigg|_{t=0}\right]
\rho^{1-u}\cr
&=&
\int_0^1\upd u\,\rho^u\,A\,\rho^{1-u}\cr
&=&
[A]_\rho^\tinyK\cr
&=&
V.
\nnee
In the above calculation it is used that
\be
D(\rho||\sigma)
&=&-\Tr\rho A+\log\Tr\exp(\log\rho+A)
\label{tan:divexpr}
\ee
and $\Tr\rho A=\Tr V$. The latter vanishes by assumption.
\endproof

\begin{proposition}
If two exponential arcs $\sigma_t$ and $\tau_t$ connecting $\sigma$, respectively $\tau$
to $\rho$ have the same tangent vector at $t=0$ then they coincide.
\end{proposition}

\beginproof
Because $\sigma_t$ and $\tau_t$ have the same tangent vector at $t=0$ it follows that
\be
0&=&
\left[\log\sigma-\log\tau+D(\rho||\sigma)-D(\rho||\tau)\right]^\tinyK_\rho.
\nnee
Take the trace of this expression to find that
\be
0&=&D(\rho||\sigma)-D(\rho||\tau).
\nnee
One concludes that
\be
0&=&\left[\log\sigma-\log\tau\right]^\tinyK_\rho.
\label{tan:prop:temp}
\ee

By Proposition \ref {tan:lem} the linear map $A\mapsto [A]_\rho^\tinyK$
is invertible. Hence it is a one-to-one
map between the spaces ${\cal A}_\rho$ and $\sanull$ 
because these spaces are finite-dimensional.
From (\ref {tan:prop:temp}) it then follows that $\log\sigma-\log\tau=0$ and hence, that $\sigma=\tau$.
\endproof

The inverse $\sigma\mapsto Y_\rho(\sigma)$ 
of the exponential map $Y_\rho(\sigma) \mapsto\sigma$
could be used as a chart for the manifold $\Mo$.
This chart is affine in the case of the m-connection.
Alternatively, one can use the correspondence provided by Proposition \ref {tan:lem}
between $\sanull$ and ${\cal A}_\rho$.
It will turn out that the chart $c_\rho$ is affine in case of the e-connection.
Note that it satisfies $c_\rho(\rho)=0$.
It is said to be centered at the point $\rho$ in $\Mo$.

The transition map $c_{\rho_1}\mapsto c_{\rho_2}$
from reference point $\rho_1$ to any other reference point $\rho_2$
is given by
\be
c_{\rho_2}(\sigma)
&=&
c_{\rho_1}(\sigma)
+\log\rho_1-\log\rho_2
+D(\rho_2||\sigma)-D(\rho_1||\sigma).
\nnee
The expression in the r.h.s.~is Fr\'echet-differentiable for any density matrix $\sigma$ in $\Mo$.
One concludes that the different charts are mutually compatible.

\section{The metric}
\label{sect:met}

Eguchi \cite{ES92} shows how to derive a metric on the tangent planes starting from a diver\-gence
function. The obvious diver\-gence function here is Umegaki's relative entropy (\ref {2:relent:def})
discussed in Section \ref {sect:arc}.

\begin{proposition}
\label{met:prop}
An inner product is defined on 
the tangent plane $T_\rho\Mo$ by
\be
(Y(\sigma),Y(\tau))_\rho
&=&
-\frac{\upd\,}{\upd s}\frac{\upd\,}{\upd t}D(\sigma_s||\tau_t)\bigg|_{s=t=0},
\nnee
where $\sigma_t$ and $\tau_t$ are exponential arcs connecting density matrices $\sigma$ and $\tau$
to the density matrix $\rho$
and $Y_\rho(\sigma)$  and $Y_\rho(\tau)$ are the tangents of $t\mapsto\sigma_t$
respectively $t\mapsto\rho_t$ at $t=0$.
The inner product is given in terms of the chart $c_\rho$ by
\be
(Y(\sigma),Y(\tau))_\rho
=
\Tr Y_\rho(\sigma)\,c_\rho(\tau)
=
\int_0^1\upd u\,\Tr\rho^u c_\rho(\sigma)\rho^{1-u}c_\rho(\tau).
\label{met:inprod}
\ee
\end{proposition}

\beginproof
One calculates
\be
\frac{\upd\,}{\upd t}\bigg|_{t=0}D(\sigma_s||\tau_t)
&=&
-\frac{\upd\,}{\upd t}\bigg|_{t=0}\Tr\sigma_s\log\tau_t\cr
&=&
-\Tr\sigma_s
\left[\log\tau-\log\rho\right] -D(\rho||\tau).
\label{2:metric:temp}
\ee
This implies
\be
(Y(\sigma),Y(\tau))_\rho
&=&
\frac{\upd\,}{\upd s}\bigg|_{s=0}
\Tr\sigma_s
\left[\log\tau-\log\rho\right]\cr
&=&
\Tr Y_\rho(\sigma) \,
\left[\log\tau-\log\rho\right]\cr
&=&
\Tr Y_\rho(\sigma)\,\left[c_\rho(\tau)+D(\rho||\tau)\right]\cr
&=&
\Tr Y_\rho(\sigma)\,c_\rho(\tau).
\nnee
The tangent vector $Y(\sigma)$ can be expressed in terms of the chart $c(\sigma)$.
This gives
\be
(Y(\sigma),Y(\tau))_\rho
&=&
\Tr\left[ c_\rho(\sigma)\right]^\tinyK_\rho
\,c_\rho(\tau).
\nnee
This can be written as (\ref {met:inprod}) because $\Tr\rho \,c_\rho(\sigma)=0$.

Let us verify that (\ref {met:inprod}) defines a non-degenerate inner product
on the tangent space $T_\rho\Mo$.

Bilinearity follows because the relation between tangent vector and chart
is linear. Positivity follows from
\be
(Y(\sigma),Y(\sigma))_\rho
&=&
\int_0^1\upd u\,\Tr\left(\rho^{(1-u)/2}c_\rho(\sigma)\rho^{u/2}\right)^\dagger
\,\left(\rho^{(1-u)/2}c_\rho(\sigma)\rho^{u/2}\right).
\nnee
Finally, $(Y(\sigma),Y(\sigma))_\rho=0$ implies that
$\rho^{(1-u)/2}c_\rho(\sigma)\rho^{u/2}=0$ for all $u$ in $[0,1]$.
This implies $c_\rho(\sigma)=0$. The latter is only possible when $\sigma=\rho$.
\endproof

Expression (\ref {met:inprod}) is Bogoliubov's inner product \cite{NVW75,PT93,NJ18}
adapted to the present notations.

\section{The dual geometry}
\label{sect:dualgeo}

With any geometry with parallel transport $\Pi$
corresponds a dual geometry \cite{AN00}
with parallel transport $\Pi^*$
given by
\be
(\Pi(\rho_1\mapsto\rho_2)V,\Pi^*(\rho_1\mapsto\rho_2)W)_{\rho_2}&=&(V,W)_{\rho_1}.
\label{dual:dualparmap}
\ee
Here $\rho_1$ and $\rho_2$ belong to the manifold $\Mo$ and $V$ and $W$ are
tangent vectors in $T_{\rho_1}\Mo$. 

A flat geometry is obtained when the parallel transport $\Pi$ is chosen
equal to the identity map, where each tangent space is
identified with the space $\sanull$ of traceless Hermitian matrices.
The geometry is that of the m-connection \cite{AN00}.
Let us verify this now.

The covariant derivative of a vector field $V$ along a smooth curve $\gamma$ is given by \cite {KMS51}
\be
[\nabla_{\dot \gamma} V]_{\gamma_t}
&=&
\frac{\upd\,}{\upd s}\bigg|_{s=0}\Pi(\gamma_{t+s}\mapsto\gamma_t)\,V(\gamma_t).
\nnee
With $\Pi$ equal to the identity map and with the path $\gamma$ given by
\be
\gamma_t&=&(1-t)\rho+t\sigma
\nnee
and the vector field given by $V(\gamma_t)=\gamma_t-\gamma_0$
one obtains
\be
[\nabla_{\dot \gamma} V]_{\gamma_t}
&=&
\frac{\upd\,}{\upd t}\gamma_t\cr
&=&
\sigma-\rho.
\nnee
The fact that the covariant derivative is constant along this path indicates that
the path is a geodesic. It is a geodesic of the m-connection.

Let us now consider the dual of the m-connection.
Because $\Pi$ is the identity map (\ref {dual:dualparmap}) simplifies to
\be
(V,\Pi^*(\rho_1\mapsto\rho_2)W)_{\rho_2}&=&(V,W)_{\rho_1}.
\label{dual:dualparmap2}
\ee
By Proposition \ref {tan:lem} there exists $A$ in ${\cal A}_{\rho_1}$ such that
$W=[A]^\tinyK_{\rho_1}$. Similarly, there exist $B$ in ${\cal A}_{\rho_2}$
such that $\Pi^*(\rho_1\mapsto\rho_2)W=[B]^\tinyK_{\rho_2}$.
From (\ref {met:inprod}) it follows that
\be
\Tr VB 
&=&(V,[B]^\tinyK_{\rho_2}\cr
&=&
(V,\Pi^*(\rho_1\mapsto\rho_2)W)_{\rho_2}\cr
&=&
(V,W)_{\rho_1}\cr
&=&
\Tr VA.
\nnee
Because $V$ is an arbitrary traceless matrix it follows that $B-A$ is a multiple of the identity $\Io$
and hence that
\be
\Pi^*(\rho_1\mapsto\rho_2)\,[A]^\tinyK_{\rho_1}&=&[A-\Tr\rho_2 A]^\tinyK_{\rho_2}.
\nnee

Choose now the vector field $V(\rho)=Y_\rho(\sigma)$
in combination with a path $\gamma$ equal to the
exponential arc $t\mapsto\sigma_t$ connecting $\sigma$ to $\rho$.
Then one finds
\be
[\nabla^*_{\dot \gamma} Y(\sigma)]_{\sigma_t}
&=&
\frac{\upd\,}{\upd s}\bigg|_{s=0}
\Pi(\sigma_{t+s}\mapsto\sigma_t)\,[c_{\sigma_{t+s}}(\sigma)]^\tinyK_{\sigma_{t+s}}\cr
&=&
\frac{\upd\,}{\upd s}\bigg|_{s=0}
[c_{\sigma_t}(\sigma)]^\tinyK_{\sigma_{t}}\cr
&=&0.
\nnee
This shows that $t\mapsto\sigma_t$ is a geodesic for the dual connection $\nabla^*$.
Because the geodesics are exponential arcs the connection is 
a non-commutative generalization of the e-connection of \cite{AN00}.

\section{The Legendre structure}
\label{sect:leg}

The relative entropy $D(\rho||\sigma)$ is convex in its first argument $\rho$.
The proof is based on Klein's inequality \cite{RD69}.
See \cite{PD08} for the more general argument based on operator monotonicity of
the function $f(x)=-x\log x$.
This convexity suggests the use of Legendre transforms.

\begin{definition}
Given a density matrix $\rho$ and a matrix $A$ in ${\cal A}_\rho$
the potential $\Phi_\rho(A)$ is defined by
\be
\Phi_\rho(A)&=&\log\Tr\exp(\log\rho+A).
\nnee
\end{definition}
It is the analogue of the logarithm of the partition sum in Statistical Physics.
The matrix $A$ corresponds with minus the Hamiltonian.
The term $\log\rho$ is added to enable that an arbitrary point of the manifold
can be taken as the center of the manifold.

Note that the Banach space of Hermitian matrices can be identified with the dual of the linear space
generated by the density matrices by
identification of the linear functional $\rho\mapsto \Tr\rho A$ with the matrix $A$ itself.
The Legendre transform of the map $\sigma\mapsto D(\sigma||\rho)$ is therefore equal to
\be
A&\mapsto&
\sup\{\Tr\sigma\, A-D(\sigma||\rho):\,\sigma\in\Mo\},
\qquad
\rho\in\Mo \mbox{ and }A\in{\cal A}_\rho.
\nnee

\begin{proposition}
For all $\rho$ in $\Mo$ and $A$ in $\sanull$ is
\be
\Phi_\rho(A)
&=&
\sup\{\Tr\sigma\, A-D(\sigma||\rho):\,\sigma\in\Mo\}.\cr
& &
\label{leg:transdef}
\ee
The maximum is reached for $\sigma=\tau_A$ with $\tau_A$ in $\Mo$ 
such that $c_\rho(\tau_A)=A$. It equals
\be
\Phi_\rho(A)
&=&D(\rho||\tau_A).
\nnee
\end{proposition}

\beginproof
From
\be
A=c_\rho(\tau_A)=\log\tau_A-\log\rho+D(\rho||\tau_A)
\nnee
one obtains
\be
\Phi_\rho(A)
&=&
\log\Tr \exp(\log\rho+A)\cr
&=&
\log\Tr\exp(\log\tau_A+D(\rho||\tau_A))\cr
&=&
D(\rho||\tau_A).
\label{leg:id}
\ee
It then follows that
\be
A&=&\log\tau_A-\log\rho+\Phi_\rho(A).
\label{dualchart:temp}
\ee
Use this to to obtain
\be
0&\le&D(\sigma||\tau_A)\cr
&=&\Tr\sigma\left[\log\sigma-\log\rho-A+\Phi_\rho(A)\right]\cr
&=&D(\sigma||\rho)-\Tr \sigma A+\Phi_\rho(A).
\nnee
This shows that for any $\sigma$ one has
\be
\Phi_\rho(A)&\ge&\Tr\sigma A-D(\sigma||\rho).
\nnee

Take now $\sigma=\tau_A$.
Then the inequality $0\le D(\sigma||\tau_A)$ in the above calculation
becomes an equality. Hence, $\sigma=\tau_A$ realizes the supremum in (\ref {leg:transdef}).
\endproof

The proposition shows that $A\mapsto \Phi_\rho(A)$ is a Legendre transform.
In particular, this implies that it is a convex function.

\section{The dual chart}
\label{sect:dualchart}

The following result is standard.

\begin{proposition}
The plane tangent to the potential $\Phi_\rho(A)$ at the contact point $\tau_A$
is the map
\be
B\mapsto \Phi_\rho(A)+\Tr\tau_A (B-A).
\nnee
\end{proposition}

\beginproof
From (\ref {leg:transdef}) one obtains
\be
\Phi_\rho(B)&\ge&\Tr\tau_AB-D(\tau_A||\rho)\cr
&=&\Tr\tau_A(B-A)+\Tr\tau_A A-D(\tau_A||\rho)\cr
&=&\Tr\tau_A(B-A)+\Phi_\rho(A).
\nnee
This shows that the plane $B\mapsto \Tr\tau_A(B-A)+\Phi_\rho(A)$ remains below the 
potential $\Phi_\rho$.
Contact at $B=A$ is clear.
\endproof

From the above proposition one concludes that the Legendre dual of the
matrix $A$ in ${\cal A}_\rho$ is the linear functional defined by the density matrix $\tau_A$.
In the approach with coordinates the derivative of the dual coordinate yields
the metric tensor: The first item of (3.32) of \cite{AN00} reads
\be
\partial\eta_j/\partial\theta^i=g_{ij}.
\nnee
The derivative of the potential gives the dual coordinate:
(3.33) of \cite{AN00} reads $\partial_i\psi=\eta_i$.
Parameter-free analogues follow below.

\begin{proposition}
Take $A$ and $B$ in ${\cal A}_\rho$.
The Fr\'echet derivative $d_{B}\,\Phi_\rho(A)$ of the potential $\Phi_\rho(A)$ in the direction $B$
equals $\Tr\tau_A B$.
\end{proposition}

\beginproof
One can write 
\be
\Phi_\rho(A+B)
&=&
\Phi_\rho(A)+\log\frac{\Tr\exp(\log\rho+A+B)}{\Tr\exp(\log\rho+A)}.
\nnee
Use now (\ref {dualchart:temp})
to obtain
\be
\Phi_\rho(A+B)
&=&
\Phi_\rho(A)+\log\Tr\exp(\log\tau_A+B)\cr
&=&
\Phi_\rho(A)+\log\left(1+\Tr[B]^\tinyK_{\tau_A}+\mbox{ o }(||B||)\right)\cr
&=&
\Phi_\rho(A)+\Tr\tau_A B+\mbox{ o }(||B||).
\nnee
\endproof

\begin{proposition}
Choose $\rho$ in $\Mo$ and $A$ and $B$ in ${\cal A}_\rho$.
The Fr\'echet derivative $d_{B}\,\tau_A$ of $\tau_A$ 
in the direction $B$ equals
\be
d_{B}\,\tau_A=\left[B-\Tr\tau_A\,B\right]^\tinyK_{\tau_A}.
\label{dual:prop:deriv}
\ee
\end{proposition}

\beginproof
One has
\be
B&=&
c_\rho(\tau_{A+B})-c_\rho(\tau_{A})\cr
&=&
\log\tau_{A+B}-\log\tau_A+D(\rho||\tau_{A+B})-D(\rho||\tau_A)\cr
&=&
\log\tau_{A+B}-\log\tau_A-\Tr\rho(\log\tau_{A+B}-\log\tau_A).
\nnee
This can be written in first order approximation as
\be
\tau_{A+B}
&=&
\exp\left(\log\tau_A+B+\Tr\rho(\log\tau_{A+B}-\log\tau_A)\right)\cr
&=&
\tau_A
+\left[B- \Tr\tau_A B\right]^\tinyK_{\tau_A}\cr
& &
+\left[\Tr\tau_A\, B+\Tr\rho(\log\tau_{A+B}-\log\tau_A)\right]\,\tau_A
+\mbox{ o }(||B||).
\nnee
Take the trace of this expression to see that the third term in the r.h.s.~vanishes.
Hence, one concludes (\ref {dual:prop:deriv}).
\endproof

\begin{proposition}
Select $\rho$ and $\sigma$ in $\Mo$ and $A$ in $\sanull$.
One has
\be
(Y(\rho),Y(\sigma))_{\tau_A}&=&\Tr c_{\tau_A}(\sigma)\,d_{B}\,\tau_A 
\nnee
with $B=c_{\tau_A}(\rho)-\Tr c_{\tau_A}(\rho)$.
\end{proposition}

\beginproof
Use $0=\Tr \tau_Ac_{\tau_A}(\rho)$ to find
$\Tr\tau_A B=-\Tr c_{\tau_A}(\rho)$ and hence $c_{\tau_A}(\rho)=B-\Tr\tau_A B$.
This is used in the now following calculation.

From Proposition \ref {met:prop} one obtains
\be
(Y(\rho),Y(\sigma))_{\tau_A}
&=&
\Tr c_{\tau_A}(\sigma)\,Y_{\tau_A}(\rho)\cr
&=&
\Tr c_{\tau_A}(\sigma)\,[c_{\tau_A}(\rho)]^\tinyK_{\tau_A}\cr
&=&
\Tr c_{\tau_A}(\sigma)\,[B-\Tr\tau_A B]^\tinyK_{\tau_A}\cr
&=&\Tr c_{\tau_A}(\sigma)\,d_{B}\tau_A.
\nnee
\endproof

\section{Affine coordinates}
\label{sect:aff}

The space $\sanull$ of traceless Hermitian matrices of dimension $n$-by-$n$ is a Hilbert space for
the Hilbert-Schmidt inner product
\be
(A,B)_\HS&=&\Tr AB,
\qquad A,B\in \sanull.
\nnee
Hence one can construct an orthonormal set $(f_i)_i$ of basis vectors in $\sanull$.
For any density matrix $\sigma$ in $\Mo$ one can write
\be
\log\sigma&=&x^i(\sigma)\, f_i+\Tr\log\sigma
\quad\mbox{ with }\quad
x^i(\sigma)=(\log\sigma,B^i)_\HS.
\nnee
The charts $c_\rho$ have vanishing expectation value. Their expansion therefore reads
\be
c_\rho(\sigma)&=&[x^i(\sigma)-x^i(\rho)]\,(f_i-\Tr\rho\, f_i).
\nnee

Introduce a field of basis vectors $e_i$ in the tangent bundle. It is defined by
\be
[e_i]_\rho&=&[f_i-\Tr\rho\,f_i]^\tinyK_\rho.
\nnee
The tangent vectors $Y(\sigma)$ can then be written as follows
\be
Y_\rho(\sigma)
=
[c_\rho(\sigma)]^\tinyK_\rho
=
[x^i(\sigma)-x^i(\rho)]\,[e_i]_\rho.
\nnee
The metric tensor $g$ is defined by
\be
g_{ij}(\rho)&=&(e_i,e_j)_\rho.
\nnee
One finds for any pair $\sigma$, $\tau$ in $\Mo$
\be
(Y(\sigma,Y(\tau))_\rho
&=&
[x^i(\sigma)-x^i(\rho)]\,g_{ij}\,[x^j(\sigma)-x^j(\rho)].
\nnee

Let us next consider the dual charts.
Take $B$ in ${\cal A}_\rho$ and expand it as
\be
B&=&B^i (f_i-\Tr\rho\,f_i).
\nnee
From (\ref {dual:prop:deriv}) one obtains for any $A$ and $B$ in $\sanull$
\be
d_B\tau_A
&=&
[B-\Tr\tau_A B]^\tinyK_{\tau_A}\cr
&=&
B^i[f_i-\Tr\tau_A f_i]^\tinyK_{\tau_A}\cr
&=&
B^i[e_i]_{\tau_A}.
\nnee
Hence one has
\be
([e_i]_{\tau_A},d_B\tau_A)_{\tau_A}=B^j(e_i,e_j)_{\tau_A}=g_{ij}B^j.
\nnee

\section{Summary and discussion}

The manifold $\Mo$ of non-degenerate $n$-by-$n$ matrices is studied in a
parameter-free way. Starting point is the notion of exponential arcs.
The tangents to such arcs span at each point $\rho$ of the manifold the
space $\sanull$ of traceless Hermitian matrices. Affine charts are introduced for both the
m- and the e-connection. The latter turn $\Mo$ into a Banach
manifold by means of a global chart $c_\rho$ centered at an arbitrary point $\rho$ of the manifold.

Bogoliubov's inner product is defined on any of the tangent planes $T_\rho\Mo$.
Parallel transport relates the different tangent planes.
The covariant derivative corresponding with the dual parallel transport is derived.
It defines the e-connection.

The divergence function is convex in its first argument. This enables the
introduction of a convex potential function $\Phi_\rho$ defined on 
the range ${\cal A}_\rho$ of the chart $c_\rho$.
The derivative of $\Phi_\rho$ defines the dual chart. 

In a final section affine coordinates are introduced. In this way the link is made
to more conventional approaches.

The identity (\ref {bogo:ident}) plays an essential role in controlling
the effects of non-commutativity. The symbol
$[A]^\tinyK_\rho$ denotes the Kubo transform of $A$ in ${\cal A}_\rho$.
See Notation \ref {not:special} in Section \ref {sect:arc}.
It maps a Hermitian matrix $A$ with vanishing expectation $\Tr\rho A=0$
onto a tangent vector with vanishing trace $\Tr [A]^\tinyK_\rho=0$.
In combination with the chart $c_\rho(\sigma)$
it allows to express the exponential map as $Y_\rho(\sigma)=[c_\rho(\sigma)]^\tinyK_\rho\mapsto\sigma$.

In the study of quantum exponential families the e-connection
can be easily derived by taking third order derivatives of the divergence function \cite{ES92}.
They yield the connection coefficients $\Gamma^k_{\,ij}$. By use of dual coordinates
it then becomes straightforward to show that exponential arcs are geodesics of a flat geometry.
In a coordinate-free approach it is more transparent to start from parallel transport.
The parallel transport of the m-connection is the identity map.
Given the metric one can then derive the dual transport and verify that the exponential arcs
are geodesics for the dual connection. This way of working is adapted here because of its
transparency.

Throughout this work the distinction is made between the spaces ${\cal A}_\rho$
of Hermitian matrices $A$ with vanishing expectation $\Tr\rho A=0$
and the space $\sanull$ of  Hermitian matrices $V$ with vanishing trace $\Tr V=0$,
although the relation between the two spaces is trivial. Doing so is clarifying.
Given a Hermitian matrix $A$
the parallel transport from $\rho_1$ in $\Mo$ to $\rho_2$ in $\Mo$ 
by means of the e-connection maps $A-\Tr\rho_1 A$
onto $A-\Tr\rho_2 A$. The commutative analogue of this transport law has been emphasized
for instance in \cite{PG13b}.

In Amari's work \cite{AS85,AN00} it is important that there is available
a potential function the Hessian of which is the metric.
It allows for an easy introduction of the Legendre duality.
It is shown in Section \ref {sect:leg} that for each $\rho$ in $\Mo$ a potential $\Phi_\rho$
can be defined on the space ${\cal A}_\rho$ in such a way that its Fr\'echet derivative,
which is the Legendre dual, is a chart affine for the m-connection.

The present paper describes the geometry of the manifold $\Mo$ of non-degenerate
$n$-by-$n$ matrices from a specific point of view. Much more is known
and the overall picture is clear.
On the other hand, the generalization to infinite dimensions consists of
separate studies such as those of \cite{SRF04a,SRF04b,JA06,CIJM19,NJ20}.
Finite-dimensional matrices are replaced by possibly unbounded operators on Hilbert space.
Density matrices are replaced by normalized positive functionals called states.
The technicality of the subject increases and many aspects
concerning the geometry of the manifold of faithful states are still unclear.

\section*{}

\end{document}